\documentclass[aps,prd,twocolumn,a4paper,superscriptaddress]{revtex4-1}
\usepackage{graphicx}
\usepackage{multirow}
\usepackage{latexsym}
\usepackage{hyperref}
\usepackage[british]{babel}
\begin{document}

\newcommand{\be}{\begin{equation}}
\newcommand{\ee}{\end{equation}}
\newcommand{\bq}{\begin{eqnarray}}
\newcommand{\eq}{\end{eqnarray}}
\newcommand{\bsq}{\begin{subequations}}
\newcommand{\esq}{\end{subequations}}
\newcommand{\bc}{\begin{center}}
\newcommand{\ec}{\end{center}}

\title{Stretching and Kibble scaling regimes for Hubble-damped defect networks}

\author{C. J. A. P. Martins}
\email{Carlos.Martins@astro.up.pt}
\affiliation{Centro de Astrof\'{\i}sica da Universidade do Porto, Rua das Estrelas, 4150-762 Porto, Portugal}
\affiliation{Instituto de Astrof\'{\i}sica e Ci\^encias do Espa\c co, CAUP, Rua das Estrelas, 4150-762 Porto, Portugal}
\author{I. Yu. Rybak}
\email[]{Ivan.Rybak@astro.up.pt}
\affiliation{Centro de Astrof\'{\i}sica da Universidade do Porto, Rua das Estrelas, 4150-762 Porto, Portugal}
\affiliation{Instituto de Astrof\'{\i}sica e Ci\^encias do Espa\c co, CAUP, Rua das Estrelas, 4150-762 Porto, Portugal}
\affiliation{Faculdade de Ci\^encias, Universidade do Porto, Rua do Campo Alegre 687, 4169-007 Porto, Portugal}
\author{A. Avgoustidis}
\email[]{Anastasios.Avgoustidis@nottingham.ac.uk}
\affiliation{School of Physics and Astronomy, University of Nottingham, University Park, Nottingham NG7 2RD, United Kingdom}
\author{E. P. S. Shellard}
\email[]{E.P.S.Shellard@damtp.cam.ac.uk}
\affiliation{Centre for Theoretical Cosmology, Department of Applied Mathematics and Theoretical Physics, University of Cambridge, Wilberforce Road, Cambridge CB3 0WA, United Kingdom}

\date{11 November 2016}

\begin{abstract}
The cosmological evolution of topological defect networks can broadly be divided into two stages. At early times they are friction-dominated due to particle scattering and therefore non-relativistic, and may either be conformally stretched or evolve in the Kibble regime. At late times they are relativistic and evolve in the well known linear scaling regime. In this work we show that a sufficiently large Hubble damping (that is a sufficiently fast expansion rate) leads to a linear scaling regime where the network is non-relativistic. This is therefore another realization of a Kibble scaling regime, and also has a conformal stretching regime counterpart which we characterize for the first time. We describe these regimes using analytic arguments in the context of the velocity-dependent one-scale model, and we confirm them using high-resolution $4096^3$ field theory simulations of domain wall networks. We also use these simulations to improve the calibration of this analytic model for the case of domain walls.
\end{abstract}
\maketitle

\section{Introduction}

A wide range of physical phenomena relies on the concept of symmetry breaking. Condensed matter and particle physics are good examples of contexts where this mechanism has been fruitfully applied. Cosmology, having deep connections with high energy phenomena in particle physics, shares this tendency.  In the cosmological context, the full implications of symmetry breakings were first realized by Kibble~\cite{Kibble}.  In particular, the result of a parity (or generically any discrete) symmetry breaking in a cosmological framework is the formation of domain wall structures, whose presence is tightly constrained by observational reasons: the current two-sigma upper bound on their symmetry breaking scale, coming from a full Markov Chain Monte Carlo analysis which marginalizes over the relevant cosmological parameters, is $0.93$ MeV \cite{Lazanu}. Several weaker constraints, coming from more qualitative analyses or invoking additional simplifying assumptions, have also been obtained in \cite{Zeldovich,Friedland,Conversi,Battye,Sousa}.

In addition to the opportunity to use these upper bounds to constrain underlying particle physics models, there are further motivations for the presence of light domain walls, a recent one being in the context of the symmetron screening mechanism for modified theories of gravity~\cite{Mota,Pogosian}. Last but not least, the scaling behavior of domain wall networks has key similarities with that of cosmic string and superstring networks, which in turn are motivated by a number of fundamental theories of the early Universe, including brane inflation scenarios~\cite{SarangiTye,PolchinskiCopelandMyers} and supersymmetric grand unified theories (GUT)~\cite{JeannerotRocherSakellariadou}. Since domain wall networks can be simulated with better resolution than the cosmic string networks and since both of them have similar properties, we can use the former to improve our understanding of the evolution of the latter.

The broad features of the cosmological evolution of topological defect networks are well understood, and have been extensively described \cite{ShellardVilenkin}. They can be quantitatively described using the velocity-dependent one-scale (VOS) model \cite{MartinsShellard,MartinsShellard2,VOSbook}. Broadly speaking, the evolution can be divided into two stages. At early times the networks are friction-dominated due to particle scattering and therefore non-relativistic. In this regime there are two possibilities for their evolution: either they are conformally stretched (in which case energy losses are negligible) or they evolve in the Kibble regime \cite{KibbleA,KibbleB}. At late times friction due to particle scattering becomes negligible and the networks become relativistic and evolve in the well known linear scaling regime. The attractor nature of this solution, and the fact that in this case the scaling velocity is a substantial fraction of the speed of light are well established, for example in the radiation and matter eras.

In this work we revisit the issue of the role of damping mechanisms on the evolution of defect networks. In particular, we show that a sufficiently large Hubble damping (specifically, a power-law dependence of the scale factor $a\propto t^\lambda$ with $\lambda$ below but close to unity) leads to a linear scaling regime where the network is non-relativistic. This is therefore another realization of a Kibble scaling regime, allowing us to interpret the linear scaling solution as the Kibble regime for Hubble-damped networks. Moreover, there is also a stretching regime counterpart for Hubble-damped networks, whose behavior we first derive analytically in a VOS model context and then confirm numerically.

The numerical part of the present work can be considered as an extension of previous results~\cite{LeiteMartins,LeiteMartinsShellard,MartinsShellardAvgoustidis}. We carry out high-resolution $4096^3$ field theory simulations of domain wall networks in Friedmann-Lema\^{\i}tre-Robertson-Walker (FLRW) backgrounds for very large expansion rates, between $\lambda=0.97$ and $\lambda=0.9999999$. This allows us to numerically confirm the presence of the aforementioned stretching and Kibble (non relativistic linear) scaling regimes for these networks.

The extended range of the expansion rates simulated yields an additional benefit. The low velocities and high densities of these fast expansion networks imply that the averages of sets of simulations with the same expansion rate but different initial conditions will have smaller statistical uncertainties than those for slower expanding boxes of the same size. Additionally, we have carried out numerical tests allowing us to estimate the systematic uncertainties in the measurements of the numerical quantities relevant for the calibration of the model. Both of these allow us to further improve the calibration of the VOS model for domain walls that was presented in \cite{MartinsShellardAvgoustidis}. Last but not least, we also carry out a consistency test of the model by comparing the model parameters relevant for the stretching regime with those obtained in the linear scaling one.

\section{Scaling regimes} \label{modelling}

The velocity-dependent one-scale (VOS) model provides a quantitative thermodynamic description of the evolution of a network of topological defects, based on a lenghtscale $L$ (to be thought of as the typical separation, correlation length or curvature radius of the defects) and a root-mean squared velocity $v$. The evolution equations for both quantities are \cite{VOSbook}
\be
(4-n)\frac{dL}{dt}=(4-n)HL+v^2\frac{L}{\ell_d}+cv\,,
\ee
\be
\frac{dv}{dt}=(1-v^2)\left(\frac{k}{L}-\frac{v}{\ell_d}\right)\,.
\ee
Here $n$ is the dimension of the defect worldsheet ($n=1,2,3$ for monopoles, cosmic strings and domain walls respectively), $H$ is the Hubble parameter, $\ell_d$ is a damping length including the effects of expansion and friction due to particle scattering
\be
\frac{1}{\ell_d}=nH+\frac{1}{\ell_f}\,,
\ee
with the friction length given by
\be
\ell_f\propto a^{1+n}\,,
\ee
and finally $c$ and $k$ are phenomenological parameters to be calibrated against numerical simulations. Note that as a first approximation these parameters may be taken as constants, but this assumption needs to be relaxed if one is interested in more accurate modeling \cite{MartinsShellardAvgoustidis}.

Now, following Kibble \cite{KibbleA,KibbleB} we note that in damped regimes the velocity is expected to change slowly. Then from the velocity equation it follows that
\be
v\sim k\frac{\ell_d}{L}
\ee
and substituting this in the evolution equation for the correlation length we find
\be
(4-n)\frac{dL}{dt}=(4-n)HL+k(c+k)\frac{\ell_d}{L}\,,
\ee
We now have four possible scenarios, depending on whether the damping is provided by friction or by the expansion of the universe, and on whether or not the velocities are negligible. 

Starting with the case where the damping length is due to friction, $\ell_d=\ell_f$, as discussed in \cite{MartinsShellard,VOSwalls} there is a transient scaling solution
\be
L\propto a\,,\quad v\propto\frac{\ell_f}{a}\propto a^n\;
\ee
this is known as the stretching regime, and corresponds to the case where the network's average velocity and density are sufficiently small to make the probability of self-intersections negligible. In this case the network is conformally stretched by the expansion. However, velocities are increasing so this regime must be a transient, at least in an expanding universe. If friction domination persists, eventually these assumptions will no longer hold, and the network will switch to the canonical Kibble regime  \cite{KibbleA,KibbleB}
\be
L\propto \left(\ell_f t\right)^{1/2}\,,\quad v\propto \left(\frac{\ell_f}{t} \right)^{1/2};
\ee
here the energy losses are significant, and therefore the correlation length grows faster than in the stretching regime while the velocity grows more slowly.

Now let us repeat this analysis if the damping is simply provided by the expansion of the universe; as we will confirm numerically in what follows, a sufficiently fast expansion rate is enough to make the defects move with non-relativistic speeds. In this case the damping length is $\ell_d^{-1}=nH$ and we will consider generic expansion rates $a\propto t^\lambda\propto \tau^{\lambda/(1-\lambda)}$, where $t$ and $\tau$ are, respectively, physical and conformal time. In this case the correlation length equation becomes
\be
(4-n)\frac{dL}{dt}=(4-n)HL+k\frac{c+k}{nHL}\,,
\ee
and just like in the previous case we find two scaling regimes. The transient scaling regime also corresponds to conformal stretching
\be
L\propto a\,,\quad v\propto\frac{1}{n\lambda}\frac{t}{a}  \propto \frac{1-\lambda}{n\lambda}\,\tau \,.
\ee
On the other hand, the analogue of the Kibble regime is precisely the standard linear scaling regime,
\be
L=\sqrt{\frac{k(c+k)}{(4-n)n\lambda(1-\lambda)}} t\,,\quad v=\sqrt{\frac{4-n}{n}\frac{1-\lambda}{\lambda}\frac{k}{k+c}}\,.
\ee

The two friction-dominated regimes and the linear scaling regime are all well known, but in what follows we will use high-resolution field-theory simulations to study the Hubble-damped stretching regime, as well as to confirm that it is a transient which eventually switches to the linear regime.

\begin{figure}[!]
\begin{center}
\includegraphics[width=3.5in]{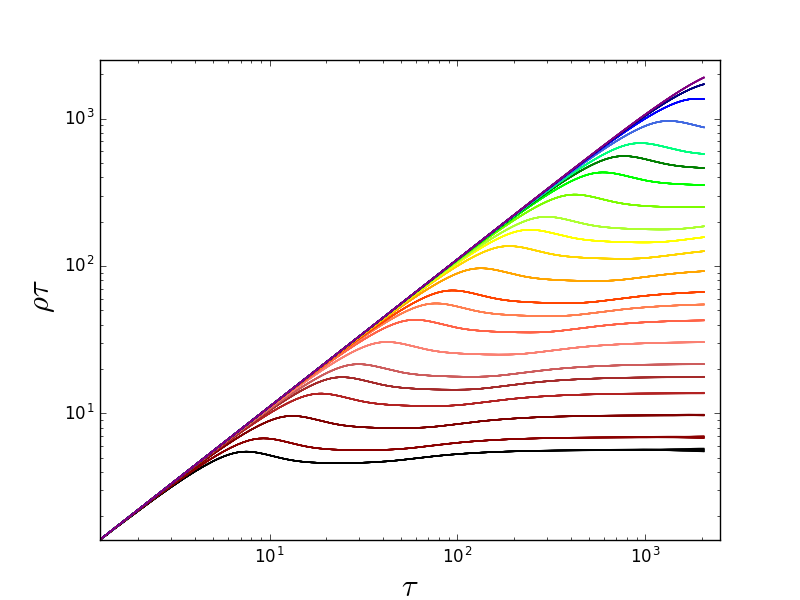}
\includegraphics[width=3.5in]{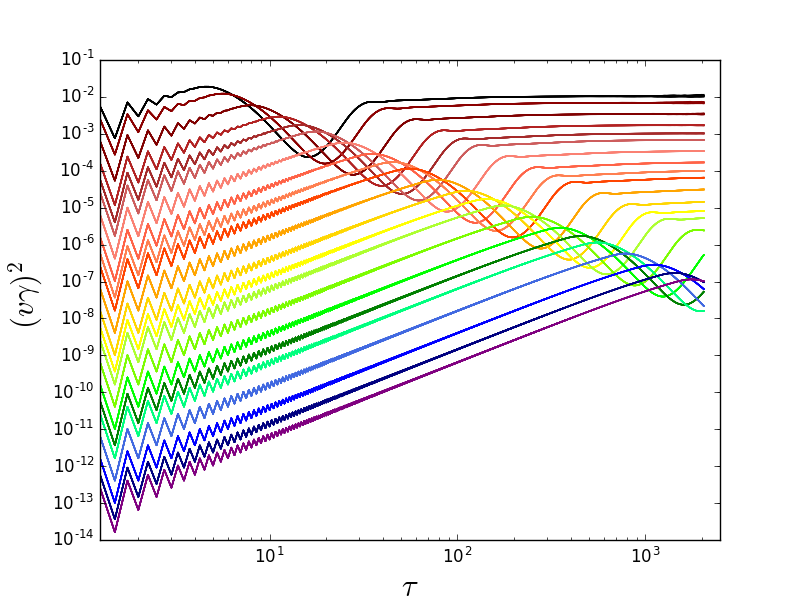}
\end{center}
\caption{\label{lambdaboxes}The evolution of the dimensionless density ($\rho\tau$, top panel) and the rms speed $(\gamma v)^2$ (where $\gamma$ is the Lorentz factor, bottom panel) in $4096^3$ domain wall simulations with different expansion rates, from $(1-\lambda)=0.03$ to $(1-\lambda)=10^{-7} $ (from black-red to blue-purple). Each line is the average of the 10 simulations, with random initial conditions.}
\end{figure}

\section{Field theory simulations} \label{Simulation}

Our numerical simulations follow in the footsteps of those reported in previous works~\cite{LeiteMartins,LeiteMartinsShellard,MartinsShellardAvgoustidis}. We assume a domain wall producing field theory model with a single scalar field $\phi$, in Friedmann-Lemaitre-Robertson-Walker universes with power law expansion rates, $a\propto t^\lambda$. After using the Press, Ryden and Spergel procedure~\cite{PRS} the equation of motion in conformal time $\tau$ has the form
\begin{equation}
   \label{Field-theoryEq}
   \frac{\partial^2 \phi}{\partial \tau^2} + 3 \frac{\mathrm{d} \ln a}{\mathrm{d} \ln \tau} \frac{\partial \phi}{\partial \tau} - \frac{\partial^2 \phi}{\partial x^i \partial x_i} = - \frac{\partial V}{\partial \phi}\,,
\end{equation}
where the potential has the form $V=V_0 \left( \frac{\phi^2}{\phi^2_0} - 1 \right)^2$.


\begin{table*}[ht]
\centering
\caption{Scaling properties of numerical simulations for domain wall networks with different expansion rates $\lambda$ in the linear scaling regime. In addition to the slope parameters $\mu$ and $\nu$ we also list the asymptotic values of the dimensionless density and the wall energy, as well as the range of conformal times used in each set of simulations. One-sigma statistical uncertainties are quoted throughout.}
\label{TableMeasure}
\begin{tabular}{c | c c | c c | c }
\hline
$\lambda$ & $\mu$ & $\nu$ & $\xi_c/\tau$ & $\gamma v$ & Fit range ($\tau$)  \\
\hline
0.97 & $-0.993 \pm 0.001$ & $\, 0.006 \pm 0.001$ & $0.177 \pm 0.001$ & $0.102 \pm 0.001$ & $500-1500 $ \\
0.98 & $-0.992 \pm 0.001$ & $\, 0.005 \pm 0.001$ & $0.145 \pm 0.001$ & $0.083 \pm 0.001$ & $500-1500 $ \\
0.99 & $-0.990 \pm 0.0003$ & $\, 0.004 \pm 0.001$ & $0.103 \pm 0.0004$ & $0.059 \pm 0.0002$ & $500-1500 $ \\
0.995 & $-0.992 \pm 0.0003$ & $\, 0.001 \pm 0.001$ & $0.073 \pm 0.0002$ & $0.041 \pm 0.0002$ & $900-2046 $ \\
0.997 & $-0.991 \pm 0.0002$ & $\, 0.006 \pm 0.0004$ & $0.057 \pm 0.0001$ & $0.032 \pm 0.0001$ & $1050-2046 $ \\
0.998 & $-0.989 \pm 0.0002$ & $\, 0.013 \pm 0.0002$ & $0.046 \pm 0.0001$ & $0.026 \pm 0.0001$ & $1200-2046 $ \\
0.999 & $-0.982 \pm 0.0001$ & $\, 0.018 \pm 0.0002$ & $0.033 \pm 0.0001$ & $0.018 \pm 0.0001$ & $1350-2046 $ \\
0.9995 & $-0.971 \pm 0.0001$ & $\, 0.031 \pm 0.0001$ & $0.023 \pm 0.0001$ & $0.013 \pm 4 \cdot 10^{-5}$ & $1500-2046 $ \\
0.9997 & $-0.960 \pm 0.0001$ & $\, 0.044 \pm 0.0001$ & $0.018 \pm 0.0001$ & $0.010 \pm 3 \cdot 10^{-5}$ & $1650-2046 $ \\
0.9998 & $-0.949 \pm 0.00003$ & $\, 0.055 \pm 0.0001$ & $0.015 \pm 3 \cdot 10^{-5}$ & $0.008 \pm 2 \cdot 10^{-5}$ & $1800-2046 $ \\

\hline
\end{tabular}
\end{table*}

Numerically the fields are rescaled so that the minima of the potential are at $\phi_0 = \pm1$, while the local maximum of the potential is at $\phi=0$ is $V_0 = \pi^2 / 2 W_0^2$ (where $W_0=10$ is the initial wall thickness in grid units). Key features of the simulation are as follows
\begin{itemize}
\item We use the WALLS code \cite{LeiteMartins,LeiteMartinsShellard}, optimized for the Intel Xeon Phi architecture and run on the COSMOS supercomputer.
\item The box size of the simulations is $4096^3$. Each simulation starts with $\tau_i=1$ and is stopped at $\tau_f=2048$ - when the horizon becomes half the box size, ensuring that the periodic boundary conditions of the simulation boxes do not affect the results. Each such simulation requires 1 Tb of memory and takes about 4 hours of wall clock time to run on 512 CPUs. The exact time required depends on $\lambda$, with faster expansion rates requiring more time: the reason is that these are denser networks, and therefore the part of the code that determines the wall densities and velocities takes longer.
\item We explored a range of fast expansion rates, with $\left( 1-\lambda \right)$ spanning the range $\left[0.03-10^{-7} \right]$. These complement the simulations reported in \cite{MartinsShellardAvgoustidis}, where we reported on simulations of slower expansion rates $0.1\le\lambda\le0.95$. For each choice of expansion rate we have carried out 10 different simulations with random initial conditions, and the results we present correspond to the average of each set of 10 runs.
\end{itemize}
Directly measured parameters from the simulations are a dimensionless density $\rho \tau =0.889 \tau / \xi_c$ and a kinetic energy $(\gamma \upsilon)^2$, where $\xi_c$ is the conformal correlation length (the numerical factor of $\sqrt{2}\pi/5\sim0.889$ stems from our choices of parameters for the potential and the thickness of the domain walls), $\upsilon$ is root mean square velocity of the network and $\gamma$ is Lorentz factor.

Figure \ref{lambdaboxes} depicts these diagnostics for all the expansion rates we have simulated. The different scaling regimes at early and late times are clearly visible, as is the fact that the transition between the early (transient) and the late (attractor) one occurs later for faster expansion rates. The oscillations in the convergence to the attractor solution, which were already noticed and discussed in \cite{LeiteMartins,LeiteMartinsShellard,MartinsShellardAvgoustidis}, are mainly due to our (numerical) choice of initial conditions.

In this section we are mainly interested in the linear scaling regime of the network. Visual inspection of Fig.  \ref{lambdaboxes} suggests that the networks reach this in the final part of the simulation, and that the faster the expansion rate the longer it takes for this regime to be reached. To accurately characterize this regime, for the purposes of further calibrating the VOS model, we need to find the interval for each $\lambda$ where the network has already reached linear scaling behavior and still possesses enough walls to enable good statistics. The last factor is a concern only for comparatively slow expansion rates. When $\lambda$ is close to unity, and the rms velocity $\upsilon$ of walls is small, the network does not lose much energy. Hence, even at the very end of simulations there are enough walls for reliable statistics. As in our previous work \cite{MartinsShellardAvgoustidis} let us introduce the scaling diagnostic parameters $\mu$ and $\nu$, defined as
\begin{equation}
   \label{Scalingmu}
   \frac{1}{\xi_c} \propto \tau^{\mu}
\end{equation}
\begin{equation}
   \label{Scalingnu}
   \gamma v \propto \tau^{\nu}\,,
\end{equation}
where for the scaling network $\mu = -1$ and $\nu = 0$. Table~\ref{TableMeasure} shows the results of this analysis, confirming that the linear scaling regime is clearly reached for the (comparatively) slower of our expansion rates. Note that in all these cases the network is scaling with a non-relativistic speed. One can therefore think of this as a Kibble regime, in the same physical sense originally described in \cite{KibbleA,KibbleB}.  On the other hand, simulations for faster expansion rates did not have enough time to fully reach the linear scaling regime---we will consider them later in this work.

\begin{figure}[!]
\begin{center}
\includegraphics[width=3.0in]{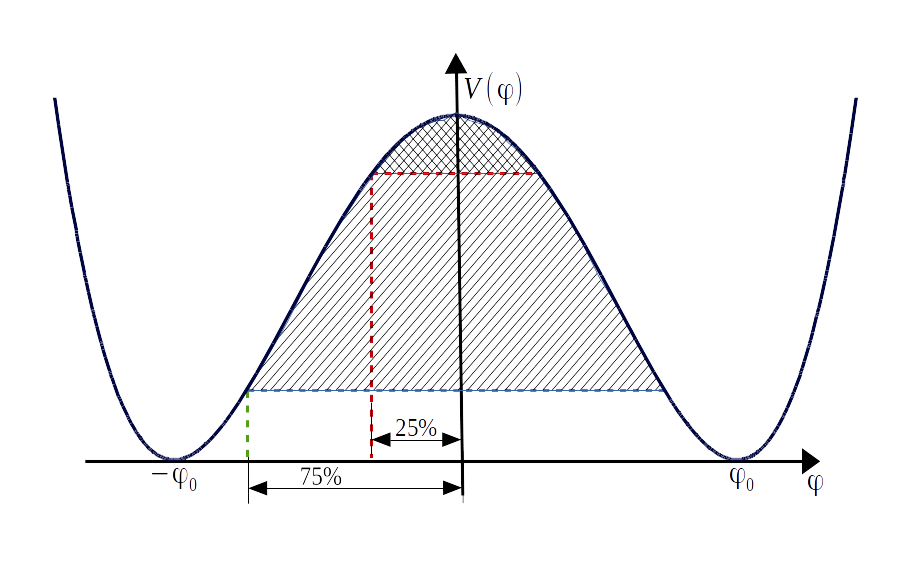}
\includegraphics[width=3.5in]{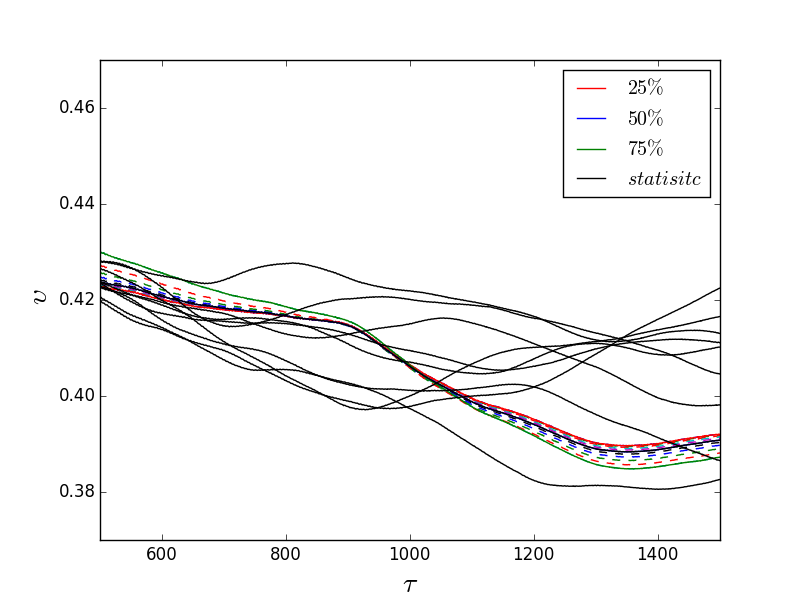}
\includegraphics[width=3.5in]{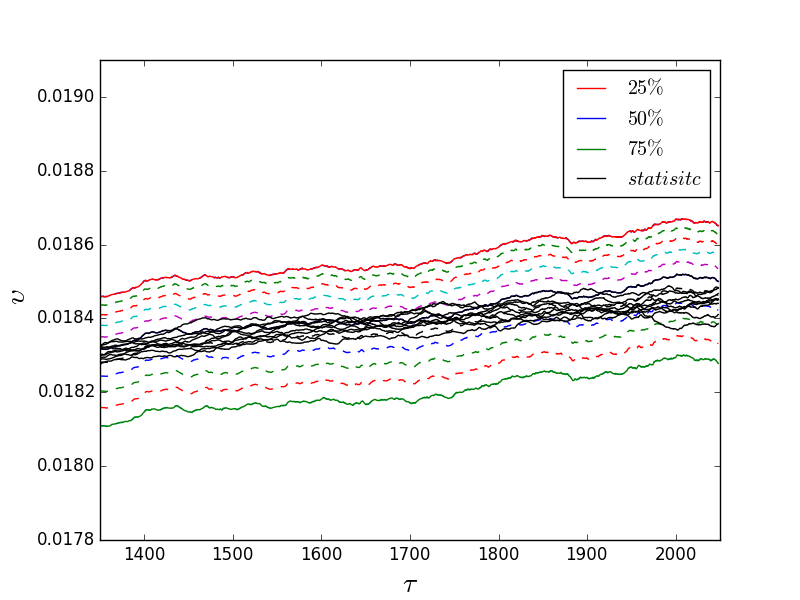}
\end{center}
\caption{\label{fig:Potential} Results from velocity measurements for different choices of thresholds, illustrated in the top panel. The middle panel presents the result for expansion rate $\lambda=0.4$ and the bottom one for $\lambda=0.999$. Black lines represent the velocities in the ten individual simulations with different initial conditions, measured with a canonical $50\%$ threshold, and provide the statistical errors. Colored lines represent the velocity in one of the ten simulations, measured using different thresholds (specifically the solid red one is for the $25 \%$ threshold and the solid green one is for $75 \%$ threshold), providing an estimate of systematic uncertainties.}
\end{figure}

From table~\ref{TableMeasure} it is also noteworthy that the statistical error bars, which come from averages of ten simulations with different (random) initial conditions, decrease significantly as the expansion rate increases. Therefore, we can anticipate that systematic errors intrinsic to the simulations (having to do with the discretization, the PRS algorithm, the identification of the domain walls in the box and the estimation of their velocities) will eventually dominate the error budget for sufficiently large expansion rates. It is therefore important to obtain estimates of these systematic uncertainties.

In order to do this, we performed additional sets of simulations for different choices of threshold for velocity measurements. In the previously described simulations, the velocity is estimated as an average over all points in the box in which the field obeys $|\phi|<0.5$ (recall that the minima of the potential are at $\phi=\pm1$). This has been previously shown to be an optimal choice. As a test, we have carried out further simulations for expansion rates $\lambda=0.4$ and $\lambda=0.999$ with exactly the same initial conditions (in other words, a fixed seed) but with thresholds for the velocity calculation spanning the range from 0.25 to 0.75; this is illustrated in the top panel of Fig. \ref{fig:Potential}, and the results of these simulations are shown in its middle and bottom panels.

Making the plausible assumption that the effect of the choice of thresholds on the velocity measurements is a reliable proxy for the systematic numerical uncertainties, this test confirms our intuition that as the expansion rate is increased the statistical uncertainties decrease much faster than the systematic ones, and therefore the latter will dominate for sufficiently large expansion rates $\lambda$. Specifically, we find that the systematic error for $\lambda=0.4$ is
\begin{equation}
   \label{systlo}
   \delta \upsilon_{syst,\lambda=0.4} = \pm 12.3 \times 10^{-4}\,,
\end{equation}
while for $\lambda=0.999$ it is
\begin{equation}
   \label{systhi}
   \delta \upsilon_{syst,\lambda=0.999} = \pm 4.7\times 10^{-4}\,.
\end{equation}
Since the difference between systematic errors for different $\lambda$ is comparatively small, in what follows we will make a linear interpolation in order to estimate the systematic uncertainty for other expansion rates.

\section{Calibration for linear scaling regime} \label{Calibration}

We can now improve upon the calibration of the VOS model discussed in \cite{MartinsShellardAvgoustidis}. We will make use of the additional data discussed above, from the simulations that reach the linear scaling regime, and combine with the data from our previous paper. We also take into account our estimate of the systematic uncertainties. Firstly, let us recall that the physical correlation length in the VOS model and the comoving one measured in our domain wall simulations are related by $L = a \, \xi_c$. And if $ a \propto t^{\lambda}$ we have  \begin{equation}
   \label{ConformalXi}
   \frac{\xi_c}{\tau}=(1-\lambda)\frac{L}{t}=(1-\lambda)\epsilon\,.
\end{equation}
The evolution equations for the standard VOS model correspond to the particular $n=3$ case of the generic model discussed in Sect. \ref{modelling}. However, as shown in our previous work, assuming that the energy loss term $c$ and the momentum parameter $k$ (the influence of local curvature on the network dynamics) are constants is not sufficient for accurate modeling. A combination of analytic and numerical arguments in \cite{MartinsShellardAvgoustidis} thus led to an extended VOS model, whose evolution equations are
\begin{eqnarray}
   \label{AveragedEqSyst3}
   \frac{dL}{dt} &=& (1 + 3 \upsilon^2)H L + c_w \upsilon + d[k_0-k(v)]^r ,  \nonumber \\
   \frac{d\upsilon}{dt} &=& (1-\upsilon^2) \left( \frac{k(v)}{L} - 3 H \upsilon \right),
\end{eqnarray}
where
\be\label{defk}
k(v)=k_{0} \frac{1-\left( q  \upsilon^2  \right)^{\beta} }{1 + \left(q \upsilon^2 \right)^{\beta} }
\ee
and $c_w$, $d$, $q$, $k_0$, $\beta$ and $r$ are the free parameters to be numerically calibrated.

\begin{figure}[!]
\begin{center}
\includegraphics[width=3.5in]{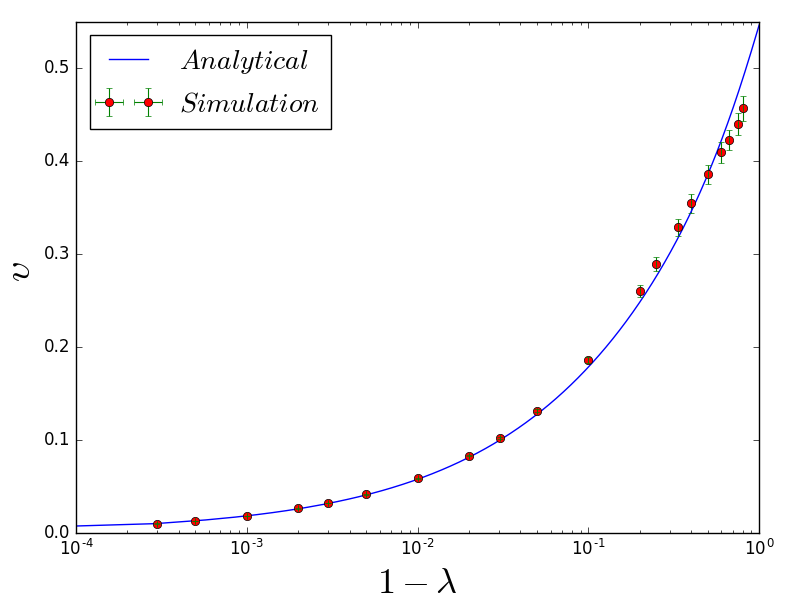}
\includegraphics[width=3.5in]{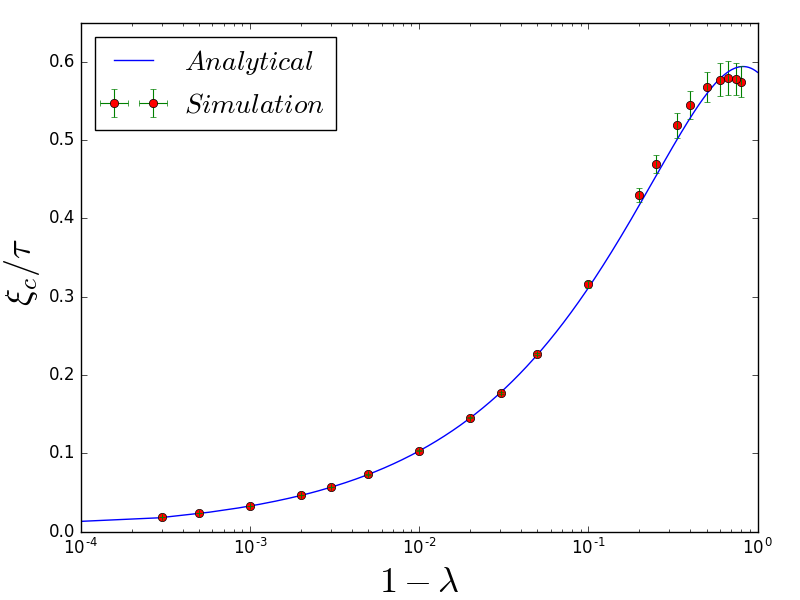}
\end{center}
\caption{\label{fig:Fitted} Velocity $v$ and conformal correlation length divided by conformal time $\xi_c/\tau$ obtained from the extended VOS model with the best-fit parameters indicated in the last line of \protect\ref{tableFits}, compared to the data from the numerical simulations for different expansion rates.}
\end{figure}

Combining the new data in Table~\ref{TableMeasure} with that in our previous work, and adding the statistical and (estimated) systematic uncertainties in quadrature for each expansion rate, we obtain the best-fit values for each of these model parameters indicated in Table \ref{tableFits}. For comparison, this table also contains the values obtained in our previous work, as well as for the case where only the fast expansion rates are used (with the caveat that this will be the case where the results of our fit will be most vulnerable to our estimates of the systematic uncertainties). A comparison between the model (with the best-fit values of each parameter) and the simulations is also shown in Fig.~\ref{fig:Fitted}.


\begin{table*}[ht]
\centering
\caption{The best-fit values for the free parameters of the extended VOS model in our present analysis; for comparison, we also show the values obtained in our previous work \protect\cite{MartinsShellardAvgoustidis}. One-sigma statistical uncertainties are quoted throughout.}
\label{tableFits}
\begin{tabular}{| c c c | c c c | c c c |}
\hline
$\lambda$ & Systematics & Reference & $c_w$ & $d$ & $r$ & $k_0$ & $q$ & $\beta$ \\
\hline
$0.5\le\lambda\le0.9$ & No & \cite{MartinsShellardAvgoustidis} & $0.00\pm0.03$ & $0.29\pm0.01$ & $1.30\pm0.06$ & $1.72\pm0.03$ & $4.10\pm0.17$ & $1.65\pm0.12$  \\
$0.2\le\lambda\le0.95$ & No & \cite{MartinsShellardAvgoustidis} & $0.00\pm0.01$ & $0.28\pm0.01$ & $1.30\pm0.02$ & $1.73\pm0.01$ & $4.27\pm0.10$ & $1.69\pm0.08$  \\
\hline
$0.97\le\lambda\le0.9998$ & Yes & This work & $0.01\pm0.01$ & $0.10\pm0.09$ & $1.34\pm0.25$ & $1.82\pm0.02$ & $1.20\pm0.72$ & $0.94\pm0.38$  \\
$0.2\le\lambda\le0.9998$ & Yes & This work & $0.00\pm0.08$ & $0.26\pm0.02$ & $1.42\pm0.04$ & $1.77\pm0.03$ & $3.35\pm0.32$ & $1.08\pm0.07$  \\
\hline
\end{tabular}
\end{table*}

It is worthy of note that the loop chopping efficiency is still statistically consistent with zero, with the exception of the case where only expansion rates $\lambda\ge0.97$ are used in the fit in which case a non-zero value is preferred at one sigma. This is in agreement with our expectations, already discussed in \cite{MartinsShellardAvgoustidis}, and confirms the expectation that the production of 'wall blobs' is a subdominant energy loss mechanism, unlike the analogous production of loops in cosmic string networks. As for the other fitted parameters, we find general agreement with our previous analysis: the only parameter that is significantly changed is $\beta$, which leads to a corresponding shift (and an increased error bar) for $q$, with which it is clearly correlated. Shifts in the remaining parameters are within about one standard deviation.

It is particularly suggestive that a value of $\beta=1$ provides a good fit to the simulations. In this case the momentum parameter has the simpler form
\be\label{defk2}
k(v)=k_{0} \frac{1-q  \upsilon^2}{1 + q \upsilon^2 }\,,
\ee
to be contrasted with the analogous parameter for cosmic strings
\be\label{defk3}
k_{s}(v)\propto k_{0,s} \frac{1-8\upsilon^6}{1 + 8 \upsilon^6 }\,.
\ee
It is tempting to speculate that the velocity dependence in the former would be associated with energy losses due to scalar radiation \cite{VEV}, which are indeed dominant in our simulations from which this calibration has emerged. Similarly, the velocity dependence in the latter could be associated with quadrupole radiation---which is implicitly assumed as an energy loss mechanism in the phenomenological modeling which, together with a comparison with Goto-Nambu string simulations, led to the analogous expression for strings \cite{MartinsShellard2}. Exploring this hypothesis is beyond the scope of the present paper, but it is certainly worthy of future study.

\begin{table*}[ht]
\centering
\caption{Scaling properties of numerical simulations for domain wall networks with different expansion rates $\lambda$ in the conformal stretching regime. In addition to the slope parameters $\mu$ and $\nu$ we also list the asymptotic values of the (constant) comoving correlation length and the slope of the wall energy, as well as the range of conformal times used for the fits in each set of simulations. One-sigma statistical uncertainties are quoted throughout.}
\label{TableDamp}
\begin{tabular}{c | c c | c c | c }
\hline
$1-\lambda$ & $\mu$ & $\nu$ & $\xi_c$ & $\gamma v/ \tau (10^{-6})$ & Fit range ($\tau$)  \\
\hline
$3\cdot10^{-6}$ & $-0.001 \pm 0.001$ & $\, 0.997 \pm 0.217$ & $0.900 \pm 0.001$ & $3.792 \pm 0.357$ & $1-76 $ \\
$2\cdot10^{-6}$ & $-0.005 \pm 0.003$ & $\, 0.967 \pm 0.125$ & $0.905 \pm 0.006$ & $2.502 \pm 0.142$ & $1-226 $ \\
$1\cdot10^{-6}$ & $-0.007 \pm 0.004$ & $\, 0.955 \pm 0.099$ & $0.908 \pm 0.008$ & $1.245 \pm 0.058$ & $1-376 $ \\
$5\cdot10^{-7}$ & $-0.007 \pm 0.003$ & $\, 0.956 \pm 0.085$ & $0.908 \pm 0.008$ & $0.623 \pm 0.025$ & $1-526 $ \\
$3\cdot10^{-7}$ & $-0.007 \pm 0.003$ & $\, 0.956 \pm 0.075$ & $0.908 \pm 0.008$ & $0.374 \pm 0.014$ & $1-676 $ \\
$2\cdot10^{-7}$ & $-0.006 \pm 0.003$ & $\, 0.956 \pm 0.068$ & $0.907 \pm 0.008$ & $0.249 \pm 0.009$ & $1-826 $ \\
$1\cdot10^{-7}$ & $-0.004 \pm 0.002$ & $\, 0.968 \pm 0.061$ & $0.905 \pm 0.005$ & $0.125 \pm 0.004$ & $1-976 $ \\
\hline
\end{tabular}
\end{table*}

\section{The conformal stretching regime}\label{stretch}

We now turn our attention to the early part of our field theory simulations. The analysis of Sect. \ref{modelling} leads us to expect a conformal stretching regime which translated into the numerically measured quantities should correspond to
\be
\rho_w\propto const. \,,\quad v\propto \frac{1-\lambda}{\lambda}\,\tau\,;
\ee
naturally the Lorentz factor is irrelevant for non-relativistic speeds. The proportionality factor can be more quantitatively calculated from the VOS model equations themselves, cf. Eqs. (\ref{AveragedEqSyst3}), and we find
\be\label{confscal}
v= \frac{15}{2\pi}\frac{1-\lambda}{1+2\lambda}\,k_0\rho_w\tau\,,
\ee
where we have used our definition of $k(v)$ in Eq. (\ref{defk})---which in the non-relativistic limit is simply $k_0$---and again the numerical factor stems from our choices of parameters for the potential and the thickness of the domain walls.

A simple visual inspection of Fig. \ref{lambdaboxes} already indicates that such a transient regime indeed occurs at early times of our simulations, and as expected it lasts longer for faster expansion rates. In order to further quantify this behavior we will again use the scaling diagnostic parameters $\mu$ and $\nu$ defined in Eqs. (\ref{Scalingmu}) and (\ref{Scalingnu}). For the conformal stretching regime we expect $\mu=0$ and $\nu=1$. Table \ref{TableDamp} shows the results of this analysis and fully confirms the presence of this regime, for the faster expansion rates for which it persists longer.

\begin{figure}
\begin{center}
\includegraphics[width=3.5in]{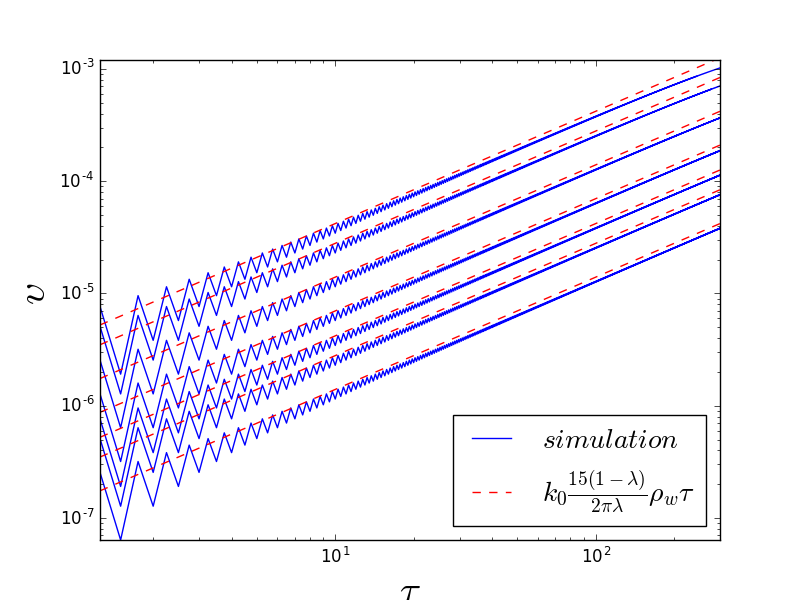}
\end{center}
\caption{\label{fig:damped}Comparing the analytic solution of the VOS model, with the free parameters fixed at their best-fit values, for domain walls in the stretching regime (red dashed lines) with the result of our field theory numerical simulations for high values of $\lambda$ (blue solid lines).}
\end{figure}

Moreover, Eq. (\ref{confscal}) also allows us to carry out a further test of the VOS model calibration, since we can use it to numerically measure the value of the parameter $k_0$. This is a useful consistency test, since in the analysis in the previous section $k_0$ as well as the other model parameters were determined using data from the linear scaling regime, and the same is true of our earlier work \cite{MartinsShellardAvgoustidis} (although in this previous work we have also shown that the calibrated model accurately described the transition from a radiation to a matter dominated universe, in which case the network is not scaling). Here we find that
\be\label{knonrel}
k_0=1.76\pm0.11\,,
\ee
which is fully consistent with the results of Table \ref{tableFits}. Figure \ref{fig:damped} shows the conformal stretching part of the evolution of our simulations, plotted together with the predictions of the VOS model with its best-fit parameters.

\section{Conclusions}

In this work we revisited the role of damping mechanisms in the cosmological evolution of topological defect networks, comparing and contrasting friction due to particle scattering and the expansion of the universe as damping mechanisms capable of making the evolution of the networks be non-relativistic. In both cases the network can either be conformally stretched or evolve in a Kibble scaling regime \cite{KibbleA,KibbleB}. In particular we have explicitly demonstrated that a sufficiently large Hubble damping (that is a sufficiently fast expansion rate) eventually leads to a linear scaling regime where the network is non-relativistic, but this is typically preceded by a stretching regime counterpart which we characterize for the first time.

Our analytic study was done in the context of the canonical VOS model \cite{VOSbook}, but also confirmed with state-of-the-art $4096^3$ field theory simulations of domain wall networks, the analysis of which is a continuation of our previous work \cite{MartinsShellardAvgoustidis}. Specifically we have for the first time simulated domain wall networks in universes with very fast expansion rates (from $1-\lambda=0.03$ to $1-\lambda=10^{-7}$, for $a\propto t^\lambda$). In addition to confirming to high accuracy the two scaling regimes, we have also used these simulations to improve the calibration of the VOS model for the case of domain walls. This improvement stems both from the wider range of expansion rates now available and from the inclusion of an estimated systematic uncertainty in our error budget.

In the future it will be interesting to exploit recent progress in computing power and code optimization and use these high-resolution simulations to gain a better understanding of the energy loss mechanisms of these networks. While for the case of domain walls our work establishes the dominance of scalar radiation, the analogous question remains unanswered for cosmic strings. In the latter case the contribution of the loop chopping term is known to be more important, but significant differences remain between the results of field theory and Goto-Nambu simulations. Although they may partially be explained by the different spatial resolution and dynamical range of both types of simulations, it is not clear that numerical differences provide a satisfactory explanation. A fully calibrated VOS model will allow a direct comparison between both types of simulations, enabling a test of their consistency. We hope to address this issue in future publications.


\begin{acknowledgments}

This work was done in the context of project PTDC/FIS/111725/2009 (FCT, Portugal). CJM is also supported by an FCT Research Professorship, contract reference IF/00064/2012, funded by FCT/MCTES (Portugal) and POPH/FSE (EC). IR is supported by an FCT fellowship (SFRH/BD/52699/2014), under the FCT PD Program PhD::SPACE (PD/00040/2012). The work of AA was supported by an Advanced Research Fellowship at the University of Nottingham, UK.

This work was undertaken on the COSMOS Shared Memory system at DAMTP, University of Cambridge operated on behalf of the STFC DiRAC HPC Facility. This equipment is funded by BIS National E-infrastructure capital grant ST/J005673/1 and STFC grants ST/H008586/1, ST/K00333X/1. The WALLS code was extensively modernised by James Briggs (COSMOS IPCC) and John Pennycook (Intel), as described in Ref. \cite{SGI}.

We respectfully dedicate this work to Tom Kibble.

\end{acknowledgments}

\bibliography{walls2}
\end{document}